\documentclass{eptcs}
\providecommand{\event}{TARK 2017} 
\usepackage{underscore}

\usepackage{version}
\includeversion{ccsin}
\includeversion{tarkin}

\usepackage{amssymb}
\usepackage{amsmath}

\usepackage{mathrsfs}

\let\citeyear\cite

\newcommand{\Circ}{\mbox{{\small $\bigcirc$}}}
\newcommand{\NCirc}{\ensuremath{\settowidth{\dimen7}{\mbox{\Circ}}%
              \makebox[0pt][l]{$\Circ$}\makebox[\dimen7]{\mbox{-}}}}
\newcommand{\NBox}{\ensuremath{\settowidth{\dimen7}{\mbox{$\Box$}}%
              \makebox[0pt][l]{$\Box$}\makebox[\dimen7]{\mbox{\raisebox{1pt}{-}}}}}
\newcommand{\NDiamond}{\ensuremath{\settowidth{\dimen7}{\mbox{$\Diamond$}}%
              \makebox[0pt][l]{$\Diamond$}\makebox[\dimen7]{\mbox{\raisebox{1pt}{\hspace{0.5pt}-}}}}}

\newcommand{\sprev}{\ensuremath{\settowidth{\dimen7}{\mbox{\Circ}}%
              \makebox[0pt][l]{$\Circ$}\makebox[\dimen7]{\mbox{s}}}}

\newcommand{\believes}{~\mathbf{believes}~}
\newcommand{\believesn}{\mathbf{believes}}

\newcommand{\sees}{~\mathbf{sees}~}
\newcommand{\seesn}{\mathbf{sees}}
\newcommand{\said}{~\mathbf{said}~}
\newcommand{\saidn}{\mathbf{said}}
\newcommand{\controls}{~\mathbf{controls}~}
\newcommand{\controlsn}{\mathbf{controls}}
\newcommand{\fresh}{\mathbf{fresh}}
\newcommand{\key}[1]{\mathrel{\stackrel{\scriptscriptstyle {#1}}{\leftrightarrow}}}
\newcommand{\pkey}[1]{\mathrel{\stackrel{\scriptscriptstyle {#1}}{\mapsto}}}

\newcommand{\cancompute}[1]{\mathit{can\_generate}_{#1}}

\newcommand{\cancomputeNF}[1]{\mathit{can\_generate}^{\scriptscriptstyle\rm
NF}_{#1}}

\newcommand{\strin}{\mathsf{s}}

\newcommand{\stringterm}{\mathbf{s}}

\newcommand{\msg}{\mathbf{m}}

\newcommand{\st}{\mathit{st}}

\renewcommand{\int}[2]{{[#1]}_{#2}}
\newcommand{\intn}[1]{[#1]}

\newcommand{\extract}[2]{\mathsf{extract}_{#1}(#2)}
\newcommand{\extractn}{\mathsf{extract}}
\newcommand{\tp}{l}

\newcommand{\good}{\mathit{good}}

 \usepackage{amsthm}

 \newtheoremstyle{theorem}{\topsep}{\topsep}%
      {\itshape}
      {}
      {\bfseries}
      {.}
      {10pt}
      {\thmname{#1}\thmnumber{ #2}\thmnote{ (#3)}}
 
 \theoremstyle{theorem}

\newtheorem{theorem}{Theorem}

\newtheorem{proposition}[theorem]{Proposition}

\newtheorem{EXAMPLE}[theorem]{Example}

\newcommand{\sat}{\models}
\newcommand{\rimp}{\Rightarrow}
\newcommand{\riff}{\Leftrightarrow}

\renewcommand{\>}{\rangle}

\renewcommand{\phi}{\varphi}

\newcommand{\cK}{\mathcal{K}}

\newcommand{\cN}{\mathcal{N}}

\newcommand{\cR}{\mathcal{R}}

\newcommand{\cT}{\mathcal{T}}

\newcommand{\R}{\mathcal{R}}
\newcommand{\I}{\mathcal{I}}
\newcommand{\K}{\mathcal{K}}

\newcommand{\C}{\mathscr{C}}

\newcommand{\sendE}[1]{\mathit{send}(#1)}
\newcommand{\recvE}[1]{\mathit{recv}(#1)}
\newcommand{\send}[2]{\mathsf{sent}_{#1}(#2)}
\newcommand{\recv}[2]{\mathsf{recv}_{#1}(#2)}

\newcommand{\fmla}[1]{#1}

\newcommand{\truep}{\mathit{true}}

\newcommand{\true}{\textbf{true}}
\newcommand{\false}{\textbf{false}}

\newcommand{\Rule}[2]{          
  \begin{array}{c}
  #1 \\\hline
  #2
  \end{array}}

\newcommand{\encr}[2]{\{#1\}_{#2}}
\renewcommand{\Pr}{\mbox{Pr}}

\newcommand{\ktrans}[1]{#1^T}

\newcommand{\mtrans}[1]{#1^M}

\newcommand{\mi}[1]{\mathit{#1}}

\newenvironment{wideitemize}[1]
   {\begin{list}{$\bullet$}
                     {\setlength{\labelwidth}{#1}
                      \setlength{\leftmargin}{#1}}}
   {\end{list}}

\newcommand{\PHI}{F}

\newcommand{\secref}[1]{Section~\ref{#1}}

\usepackage{xspace}
\newcommand{ \agpr}{it\xspace}

\begin{ccsin}
  \title{An Epistemic Foundation for Authentication Logics\\
    (Extended Abstract)}
\author{Joseph Y. Halpern
\institute{Cornell University\\
Ithaca, NY, USA}
\email{halpern@cs.cornell.edu}
\and
Ron van der Meyden
\institute{University of New South Wales\\
Sydney, Australia}
\email{meyden@cse.unsw.edu.au}
\and
Riccardo Pucella
\institute{Forrester Research\\
Cambridge, MA, USA}
\email{riccardo@acm.org}
}
\end{ccsin} 
\date{}

\def\titlerunning{An Epistemic Foundation for Authentication Logics}
\def\authorrunning{J. Y. Halpern, R. van der Meyden \& R. Pucella}

\begin{document}

\maketitle

\begin{abstract}
While there have been many attempts, 
going back to BAN logic, to base reasoning about security protocols on 
epistemic notions, they have not been all that successful. 
Arguably, this has been due to the particular logics chosen. 
We present a simple logic based on the well-understood modal operators of 
knowledge, time, and probability, and show that it is able to handle issues that have often 
been swept under the rug by other approaches, while being flexible 
enough to capture all the higher-level security notions that appear in 
BAN logic. Moreover, while still assuming that the knowledge operator
allows for  
unbounded computation, it can handle the fact that a computationally bounded 
agent cannot decrypt messages in a natural way, by distinguishing 
strings and message terms. We demonstrate that our logic can capture
BAN logic notions by  
providing a translation of the BAN operators into our logic, capturing
belief by a form of  
probabilistic knowledge.
\end{abstract}

\begin{tarkin} 
\section{Introduction}

For over 25 years now, 
there has been an intuition in the world
of security that formal theories of knowledge and belief should have
something interesting to say about  security protocols.
Many logics have been designed that embody this intuition.
One of the earliest and the most discussed is BAN logic
\cite{r:burrows90}. 
While BAN logic has been the subject of many (quite legitimate!) criticisms,
we believe that there are important features of the BAN approach that have been lost in
more recent approaches such as model checking \cite{r:lowe98,r:mitchell97},
inductive-assertions methods \cite{r:paulson98}, strand spaces
\cite{r:thayer99}, and process calculi \cite{r:gordon99}: 
namely, the ability to express intuitions of protocol
designers regarding notions such as belief, trust, freshness, and
jurisdiction.  Such high-level abstractions play a significant role in
informal reasoning about security protocols. It would be desirable 
for such intuitive ideas concerning the protocol specifications to 
be reflected in formal specifications, and for informal arguments 
concerning such notions to be reflected in formal proofs.

In this paper, we argue that a modal logic with standard notions of
knowledge, probability, and time, together with atomic predicates that
capture messages {\em sent} and {\em received} by an agent, and a
predicate that we call $\extractn$ which characterizes an agent's
ability to extract information from messages, can capture most of the
higher-level abstractions that we seem to need.  We show that such a
logic is able to handle issues that have often been swept under the
rug by other approaches, and is flexible enough to capture the
higher-level security notions that appear in BAN logic.  We do this by
providing a translation of the BAN operators into our logic, capturing
belief by a form of probabilistic knowledge, and showing that the
translation satisfies the BAN inference rules.
The translation highlights some subtleties in the BAN framework,
including some that were missed by earlier authors.

Logics in the BAN tradition have long struggled to reconcile 
the information-theoretic semantics of logics of knowledge and belief
with the computational aspects of cryptography, which raise a version of
the \emph{logical omniscience problem}. 
Suppose that $i$ sends $j$ the message $\msg'$, where $\msg'$ is
$\msg$ encrypted by a shared 
key $k$.  Does $j$ know that $i$ has sent $\msg$
encrypted by $k$?  If $j$ does 
not have the key $k$, then, intuitively, the answer is no; agent $j$ has
no way of knowing that $\msg'$ is the result encrypting $\msg$ by $k$.
Of course, if $j$ were not computationally bounded, then $j$ could
figure out that $\msg'$ is indeed $\msg$ encrypted by $k$.  Standard
approaches to modeling knowledge treat agents as computationally
unbounded; in particular, agents are assumed to know all valid formulas.
Since (given a fixed encryption framework, and assuming unique
encryptions) the fact that $\msg'$ is the result of encrypting $\msg$ by
$k$ is a valid mathematical statement, all agents will know it.  This is
the logical omniscience problem.  There have been attempts to overcome this problem: 
Cohen \cite{cohenthesis}, for instance, deals with it using what seems to us a rather
complicated 
semantics for knowledge involving permutations (see 
Section~\ref{sec:related}
for details).

We propose a simpler and arguably far more intuitive approach that 
allows us to retain the standard semantics for knowledge. 
It has been common in the literature on authentication logics to represent the 
complex message that is the result of encrypting a message $\msg$ by a
key $k$   using the {\em term} $\encr{\msg}{k}$ in both the syntax and semantics
of the logic.    
We depart from this approach by distinguishing two views of messages.
The first views a message simply as a string of symbols; the second
views the message
as 
 a term with structure.  When $j$ receives the message
$\msg'$, $j$ knows that  \agpr received (the string) $\msg'$.  What $j$ does
not know is that  \agpr received $\msg$ encrypted by $k$; $j$ considers it
possible that $\msg'$ is $\msg''$ encrypted by $k''$, or that $\msg'$
is not the encryption of any message.  To model this, we consider both
strings and terms.  What is sent or received are strings; we use the
notation $\strin = \intn{\msg}$ to denote that $\strin$ is the string
that represents the message (term) $\msg$.  We also allow for
``impossible'' runs where $\strin = \intn{\msg'}$; that is, we allow the
agent to be uncertain as to what message is represented by the string
$\strin$ (even when $\strin$ representing $\msg$ is a fact of
mathematics).  Using such impossible runs, we can easily model the fact
that $i$ may know that $\strin$ represents the encryption of some
message, even though $i$ does not know which message it is the
encryption of (in all runs that $i$ considers possible, $\strin =
\intn{\encr{\msg'}{k'}}$ for some message $\msg'$ and key $k'$) or that
$i$ knows that encryptions are unique, or that $\strin$ represents the
encryption of a message of length at most 20.  We believe that this
approach to dealing with logical omnisicience should be
useful beyond the context of this paper.
\end{tarkin} 

\section{A Logic for Security Properties}\label{s:logic} 

\subsection{Syntax}\label{s:syntax}

We use a modal 
logic for
reasoning about security protocols. We assume a finite set of
principals that for simplicity we 
represent
by integers,
a set $\cK$ of keys,
a set $\cN$ of nonces, a set $\cT$
of plaintexts,
and a set $\Phi$ of
(application-specific) primitive propositions.
We assume that $\cK$ contains both symmetric keys (used in shared-key
cryptography) and asymmetric keys (used in public-key 
cryptography), and that they can be distinguished. We also assume that 
keys, nonces, and plaintexts can be distinguished, so that $\cK$,
$\cN$, and $\cT$ are disjoint sets, and that encrypted messages can be 
distinguished from unencrypted messages.

Like Abadi and Tuttle \citeyear{AT91} (AT from now on)
and other BAN successors,
we assume that 
formulas can state properties of messages, and can also be used in
messages.  
Thus, we define formulas and messages
simultaneously
as follows, where we use $p$ for a generic element of $\Phi$, $i$
for a generic principal (or agent), $\msg$ for a generic message, $t$
for a generic plaintext, $k$ for a generic key, $n$ for a generic
nonce,
$\alpha$ for a generic real number in $[0,1]$,
$\stringterm$ for a generic term of type string,
$\strin$ for a generic concrete string,  
$x$ for a generic variable ranging over strings, 
and $\phi$ for a generic formula.
As usual, a concrete string is a sequence of symbols from some
alphabet $\Sigma$.   
We view messages both as strings and as terms with
structure.  When we write $\msg$, we are thinking of 
the message as a term
with structure, as is made clear in the following grammar:
\begin{eqnarray*}
  \stringterm &::= & \strin ~|~ x \\
\msg &::= & t ~|~ k ~|~ n ~|~ i ~|~ (\msg_1,\msg_2) ~|~ 
            \encr{\msg}{k} ~|~ \fmla{\phi} \\ 
\phi & ::= & p  ~|~  \send{i}{\stringterm} ~|~ \recv{i}{\stringterm} ~|~
\extract{i}{\msg} ~|~ \neg\phi ~|~ \phi_1\wedge\phi_2 ~|~  K_i\phi ~|~
\Circ\phi ~|~ \\
& & \NCirc\phi  ~|~ \Box\phi ~|~  
        \NBox\phi ~|~ \Pr_i(\phi)\ge\alpha ~|~ \exists x \, \phi  ~|~ 
 \intn{\msg} = \stringterm ~|~ \stringterm \sqsubseteq \stringterm' \,.
\end{eqnarray*}
Besides the application-specific primitive propositions, we also have
``built-in''
primitive propositions of the form $\send{i}{\stringterm}$, $\recv{i}{\stringterm}$, 
and $\extract{i}{\msg}$.
Note that agents send and receive 
strings, not message terms.
The proposition $\extract{i}{\msg}$ holds if $i$ can ``extract'' the
message  $\msg$ from 
strings
it has received (and other information at
its disposal). 
Exactly what $\extractn$ means depends on the application,
the 
capabilities of principals, and the protocol they are running.
For now, we make no assumptions, viewing it as a black box.  
(In \secref{s:dy}, we give a concrete implementation
of $\extractn$ capturing the Dolev-Yao capabilities.)

The knowledge operator $K_i\phi$ states that agent $i$ knows the fact
$\phi$. 
The temporal operator $\Circ\phi$ states that $\phi$ is true at the
next time step, while $\NCirc\phi$ states that $\phi$ was true at the
previous time step, 
if any. 
We will use the
abbreviations $\Circ^\tp\phi$ and $\NCirc^\tp\phi$ (for
$\tp\in\mathbb{N}$)
for the 
$\tp$-fold application of $\Circ$ and $\NCirc$, respectively, to $\phi$. 
The temporal operator $\Box\phi$ states that $\phi$ is true at the
current time, and all subsequent times. 
Similarly, $\NBox\phi$ states that $\phi$ is true at the current time, 
and all previous times. 
The formula 
$\Pr_i(\phi)\geq\alpha$ says that the formula $\phi$
holds with probability at least $\alpha$, according to  agent $i$.
The range of quantification is strings: the formula $\exists x\, \phi$ says that there exists a
 string $x$ such that $\phi$ holds. The construction $[\msg]= \stringterm$ says that 
 the string $\stringterm$ is the encoding of the message $\msg$.
 That is, we assume that every message is represented as a string.
We also assume that there is a pairing
function that maps pairs of strings to strings, and 
an encryption function that maps strings and keys to strings.  
We discuss this in more detail below.
 Finally, $\stringterm \sqsubseteq \stringterm' $ says that the string
 $\stringterm'$ can be constructed from $\stringterm$ and other strings
using the pairing and encryption functions described in
Section~\ref{semantics}.  
\begin{tarkin} 
We use the usual abbreviations, and write $\sprev\phi$ for $\NCirc\phi \land \neg \NCirc \false$
($\phi$ was true 
at the previous step, and there was a previous step).
\end{tarkin}

\subsection{Semantics}
\label{semantics}

A \emph{multiagent system}  \cite{r:fagin95} consists of $n$ agents
and an environment,
each of which is in some
\emph{local state} at a given point in time. 
We briefly review the relevant details here.

We assume that an agent's local
state encapsulates all the information to which the agent has
access. In the security setting, the local state of an agent might
include some initial information regarding keys, the messages  \agpr has
sent and received, and perhaps the reading of a clock. 
The \emph{environment state} describes information relevant to the analysis
that may not be in any agent's state.
A \emph{global state} has the form $(\st_e,\st_1,\ldots, \st_n)$, where
$\st_i$ is agent $i$'s state, for $i = 1, \ldots , n$, and $\st_e$ is the
environment state.
In general, 
the actual form of 
these
local states depends on the application.

\begin{tarkin} 
We define a \emph{run} to be a function from time to global states.  A
\emph{point} is a pair $(r, m)$ consisting of a run $r$ and a time
$m\in {\bf N}$.  At a point $(r, m)$, the system is in some global
state $r(m)$.  
If $r(m) = (\st_e, \st_1, \ldots , \st_n)$, then we take $r_i(m)$ to be
$\st_i$, agent $i$'s local state at the point $(r, m)$, and $r_e(m)$ to
be $\st_e$, the environment state.  We formally define a {\em system} to
consist of a set $\R$ of runs.
\end{tarkin}

For simplicity,
we restrict attention to a specific
class of systems, suited to modeling security protocols. These are
message-passing systems in which one (or more) of the agents is an
adversary with the capacity to monitor and control message
transmission.  
Messages have compositional structure, but are transmitted as strings. 
We assume that the local state of an agent 
at time $m$ 
 is a sequence of 
the form $\<e_0, e_1, \ldots , e_m\>$, where $e_0$ is the initial state
(which typically contains the keys and nonces that $i$ is initially
aware of),
and $e_i$ for $i \ge 1$ is 
a set of events of the form
$\sendE{j,\strin}$ or
$\recvE{\strin}$ where $\strin$ is a 
string
and 
$j$
is an agent.  
We assume that messages (as strings) are sent or received during a \emph{round},
where round $m$ takes place between times $m-1$ and $m$.
Event $\sendE{j,\strin}$ is in $r_i(m)$ if $i$ sends string $\strin$
in round $m$ of run $r$, intending that it be delivered to agent $j$,
while event $\recvE{\strin}$ is in $r_i(m)$ if $i$ receives string
$\strin$ in round $m$ of run $r$.  Note that the sender is not
included in $\recvE{\strin}$.  Intuitively, this is because the
receiver may not be able to determine the sender of a message it
receives.
For an event $x$, we abuse notation and  write $x\in r_i(m)$ to denote
that $x\in e_k$ for 
some $k\leq m$.  
The initial state
represents information such as the public and private keys to be used
during the run, and other values such as nonces to be used by the agent
during the run.

As we said, we distinguish between strings and message terms, and we
allow agents to be ``confused'' about what term a string represents.  
We use the initial environment state of a run to encode the relationship
between terms and strings.  Specifically, we take $r_e(0)$ to include a
collection of equations of the form $\intn{\msg} = \strin$, 
with exactly one such  equation for each message $\msg$. 
We write $\int{\msg}{r}$ to denote the string $\strin$ such that
$\intn{\msg} = \strin$ is in $r_e(0)$.  
There
may be 
some constraints on the relationship between strings and terms.  For
example, in contexts where all messages are commonly known to be encoded
by unique strings, we would require that there is no run $r$ and no
message terms $\msg \ne \msg'$ such that 
$\int{\msg}{r} = \int{\msg'}{r}$.
We now discuss some assumptions that we make for the purposes of this paper;
others are discussed in \secref{a:soundness}:
\begin{itemize}
\item Keys $k$ and their inverses $k^{-1}$ are strings,
and represent 
themselves in all runs; that is, $\int{k}{r} = k$ and $\int{k^{-1}}{r} =
k^{-1}$ for all keys $k$ and runs $r$.  Similarly, principals (agents),
plaintexts, nonces, and 
messages in the form of formulas are also represented as strings, and
represent themselves.

\item There is a pairing function on strings, so that if $\strin$,
$\strin'$ are strings, then there is another string that we denote
$(\strin,\strin')$.  
Moreover,  we assume that 
the string representing $(\msg_1,\msg_2)$ is the
pairing of the strings representing $\msg_1$ and $\msg_2$; that
is, $\int{(\msg_1,\msg_2)}{r} = (\int{\msg_1}{r},\int{\msg_2}{r})$ for
all runs $r$.
\item There is a \emph{run-dependent} encryption function on strings.
 That
 is, given a string $\strin$ and a key $k$, there is another string that
  we denote $\int{\encr{\strin}{k}}{r}$ that we can think of as the result
of encrypting $\strin$ by $k$ in run $r$.  
We do \emph{not} assume that 
$\int{\encr{\strin}{k}}{r} = \int{\encr{\strin}{k}}{r'}$ for all runs
$r$ and $r'$.  
An agent may be ``confused'' about how $\strin$ is encrypted.

\item{}
  We define
$\int{\encr{\msg}{k}}{r} = \int{\encr{\strin}{k}}{r}$ if 
$\int{\msg}{r} = \strin$.  
That is, 
agents ``understand'' that a message is encrypted by means of an
operation on the string that encodes the message.  

\item Encryption is unique: if $\int{\encr{\msg}{k}}{r}
=  \int{\encr{\msg'}{k'}}{r}$, then 
$\int{\msg}{r} = \int{\msg'}{r}$ and $k = k'$ for all runs $r$.
(This assumption is critical in 
the use of encryption as an authentication mechanism, and 
is typically assumed in the literature on authentication logics.) 
Moreover, $\int{\encr{\msg}{k}}{r}$ is distinct from any plaintext,
nonce, key, agent name, or pairing.

\end{itemize}

Because we want to reason about probabilities, we work with 
 \emph{interpreted (probabilistic) systems} 
of the form  $\I = (\R,\pi, \C, \{\mu_C\}_{C \in \C})$,
where $\R$ is a system, $\pi$ is an interpretation for the propositions in $\Phi$ that assigns truth
values to the primitive propositions at the global states,
$\C$ is a partition of the runs in $\R$ into cells, and for each cell $C \in \C$, 
$\mu_C$ is a probability distribution on the runs in  $C$.
The assumption is that agents are using a possibly randomized
protocol, while the adversary is using a protocol that combines
possibly non-probabilistic choices (such as choosing an agent to
attack) with probabilistic moves.  Cell $C$ ``factors out'' the
nonprobabilistic choices, so that, in all the runs in $C$, only
probabilistic choices are made.  This allows us to put a probability
$\mu_C$ on the runs in $C$.
We do not assume a single
probability distribution on $\R$, since that would require us to put a
probability on the possible protocols that the adversary is using.
(See \cite{HT} for further discussion of this approach.)

We restrict the possible interpretations $\pi$ 
so as
to fix a particular
interpretation for the primitive propositions $\recv{i}{\strin}$ and
$\send{i}{\strin}$.
Specifically, we require that 
\begin{itemize}
\item $\pi(r(m))(\recv{i}{\strin})=\true$ iff $\recvE{\strin}\in r_i(m)$,
\item $\pi(r(m))(\send{i}{\strin})=\true$ iff $\sendE{j,\strin}\in
  r_i(m)$, for some $j$.
\end{itemize}
Given our interpretation of extraction as a black box, we put no
constraints here on how $\pi$ interprets $\extract{i}{\msg}$;
however, we do assume that extraction is
monotonic, in the sense that
once an   
agent is able to extract a message, it will be able to extract it at all
future times:  
\begin{itemize} 
\item If $\pi(r(m))(\extract i \msg))=\true$ and $m \leq n$ then 
$\pi(r(n))(\extract i \msg)=\true$. 
 \end{itemize} 
As we would expect, 
\begin{itemize}
\item $\pi(r(m))(\intn{\msg} = \strin) = \true$ if 
  $\int{\msg}{r} = \strin$
  (i.e., if $\intn{\msg} = \strin$ is in $r_e(0)$).
\end{itemize}

Roughly speaking, the formula $\strin \sqsubseteq \strin'$
says that $\strin'$ can be constructed from
$\strin$ and other strings using pairing and encryption.  Since how
encryption works on strings is run-dependent, 
we first define, 
for each run $r$, a relation 
$\sqsubseteq_r$ on strings as the smallest reflexive
and transitive relation such that
\begin{ccsin}
$\strin \sqsubseteq_r (\strin,\strin')$, 
$\strin' \sqsubseteq_r (\strin',\strin)$, and
$\strin \sqsubseteq_r \int{\encr{\strin}{k}}{r}$.
\end{ccsin}
We now take 
\begin{itemize}
\item $\pi(r(m))(\strin \sqsubseteq \strin') = \true$ if 
$\strin \sqsubseteq_r \strin'$.  
\end{itemize}

The reader may wonder why we defined the $\sqsubseteq_r$ relation on
strings, rather than defining it 
as the subterm relation 
on
messages.
This formulation allows us to
model a situation where 
we have
$\strin \sqsubseteq_r \strin'$ because $\strin =
\int{\msg}{r}$ and $\strin' = \int{\encr{\msg}{k}}{r}$, but the agent
does not know this, since  \agpr does not realize that $\strin' =
\int{\encr{\msg}{k}}{r}$.  Thus,  \agpr considers a run $r'$ possible where
$\strin \, {\not\sqsubseteq}_{r'} \,\strin'$ (because, for example, we
may have  
$\strin' \ne \int{\encr{\msg}{k}}{r'}$ even if $\strin =
\int{\msg}{r'}$).

As usual \cite{r:fagin95,r:hintikka62}, we say that agent $i$ knows a
fact $\phi$ at a point $(r,m)$ if $\phi$ is true at all points $i$
cannot distinguish from $(r,m)$, and define $i$'s indistinguishability
relation $\sim_i$ by taking $(r,m)\sim_i(r',m')$ if $r_i(m)=r'_i(m')$.
Despite making these standard choices,
we do not suffer from the usual
logical omniscience problems, exactly because of our distinction between
strings and message terms, and the fact that we allow ``impossible''
runs where the string corresponding to a message is not the one that is
mathematically determined.

Given our assumption that $r_i(m)$ has the form $\<e_0, \ldots, e_m\>$,
where $e_0$ is $i$'s initial state 
and $e_i$ for $i > 1$ is a set of
events of the form $\sendE{j,\strin}$ or $\recvE{\strin}$ describing the
messages that 
$i$ sent and received, it follows that $\R$ is a \emph{synchronous}
system where 
agents
have \emph{perfect recall} \cite{r:fagin95}.

Considering synchronous systems with perfect recall makes it relatively
straightforward to give semantics to formulas of the form
$\Pr_i(\phi)\ge\alpha$.
But having a probability on runs does not suffice to give semantics to a
formula 
such as $\Pr_i(\phi)\ge\alpha$ at a point $(r,m)$.  To do this,
we need to go from probabilities on runs to probabilities on points.
Given a point $(r,m)$, let $C_r$ be the unique cell in $\C$ such that $r
\in C_r$.  Let $\K_i(r,m)$ denote the set of points that
agent $i$ cannot distinguish from $(r,m)$,
that is, the set $\{(r',m') ~|~ (r,m)\sim_i(r',m')\}$. 
Let $\C(r)$ be the set of points where the runs are
in $C_r$, 
that is,
$\C(r)$ consists of all the points of the form $(r',m')$ with $r' \in
C_r$.  The 
probability $\mu_{C_r}$ on the runs of cell $C_r$ induces a
probability $\mu_{r,m,i}$ on the points in $\K_i(r,m)\cap\C(r)$ in a
straightforward way.  If $U \subseteq \K_i(r,m)\cap \C(r)$, define
\[\mu_{r,m,i}(U) = \frac{\mu_{C_r}(\{r': (r',m) \in U\})}{\mu_{C_r}(\{r': (r',m) \in
\K_i(r,m)\cap\C(r)\})}.\]

The 
relation of satisfaction 
of a formula $\phi$ 
without free variables
in an interpreted system
$\I=(\cR,\pi)$ at point $(r,m)$, written
$(\I,r,m)\sat\phi$, is defined inductively,
as follows:
\begin{list}{}{\setlength\leftmargin{5pt}}
  \item[] $(\I,r,m) \sat p$ iff $\pi(r(m))(p)=\true$
\item[] $(\I,r,m) \sat \neg\phi$ iff $(\I,r,m)\not\sat\phi$
\item[] $(\I,r,m) \sat \phi_1\wedge\phi_2$ iff $(\I,r,m)\sat\phi_1$ and
$(\I,r,m)\sat\phi_2$ 
\item[] $(\I,r,m) \sat K_i\phi$ iff for all $(r',m')\sim_i(r,m)$,
$(\I,r',m')\sat\phi$
\item[] $(\I,r,m) \sat \Circ\phi$ iff $(\I,r,m+1)\sat\phi$
\item[] $(\I,r,m) \sat \NCirc\phi$ iff $m= 0 $ or $(\I,r,m-1)\sat\phi$
\item[] $(\I,r,m) \sat \Box\phi$ iff for all $m'\geq m$,
$(\I,r,m')\sat\phi$
\item[] $(\I,r,m) \sat \NBox\phi$ iff for all $m'\leq m$,
$(\I,r,m')\sat\phi$
\item[] $(\I,r,m) \sat \Pr_i(\phi)\geq\alpha$ iff
$\mu_{r,m,i}(\{(r',m')~|~(\I,r',m')\sat\phi\}\cap
\K_i(r,m)\cap\C(r))\geq\alpha$
\item[] $(\I,r,m) \sat \exists x\, \phi$ if, 
for some concrete string $\strin$, 
$(\I,r,m) \sat
\phi[\strin/x]$, where $\phi[\strin/x]$ is the result of
replacing all free occurrences of $x$ in $\phi$ by $\strin$.
\end{list}
As usual, we say that $\phi$ is \emph{valid} in $\I$, written $\I\sat\phi$,
if $(\I,r,m)\sat\phi$ for all points $(r,m)$ in $\I$.

We define the probabilistic knowledge operator $K_i^\alpha\phi$ as an
abbreviation for $K_i(\Pr_i(\phi)\ge 1-\alpha)$. This operator simply
means that, essentially, no matter which cell $C$ the agent thinks the
current point is in, the probability of $\phi$ according to $i$ in that cell is at
least $1-\alpha$.

As stated above, we do not impose any \emph{a priori} restrictions
on the interpretation of $\extract{i}{\msg}$.
 Intuitively, the interpretation
of $\extractn$ is meant to capture the capabilities of the agents to
take apart messages and construct new messages. While in principle a
principal may be able to extract $\msg_1$ from $(\msg_1,\msg_2)$,
it may not do so at a particular point in a system, because the
protocol that it is using does not break up $(\msg_1,\msg_2)$.
Similarly, whether $i$ can extract $\msg$ from $\encr{\msg}{k}$
depends in part on whether $i$ ``has'' key $k$ in some sense, $i$'s
protocol, and $i$'s computational ability.  
We provide an interpretation of $\extractn$ that captures the Dolev-Yao
adversary in 
\secref{s:ban}. 
However, we stress that many other interpretations of $\extractn$ are
possible, such as interpretations that capture
guess-and-validate adversaries \citeyear{r:lowe02}.

\section{Interpreting BAN Logic}\label{s:ban}

One of our claims is that the logic we introduced in
\secref{s:logic} is a good foundation for security protocol
logics. 
Burrows, Abadi, and Needham \citeyear{r:burrows90} developed BAN logic
from similar intuitions, taking belief as a primitive rather than
knowledge and probability.
To provide evidence for our claim, we show how we can interpret the
constructs of BAN logic by essentially rewriting them into the simpler
primitives of our
logic. 
Although we focus here on BAN logic
as an example, 
we believe that we could similarly reconstruct other related logics.

\subsection{Definition of BAN Logic}

We reformulate the syntax of BAN logic, along the lines of AT.
The set of formulas and set of messages are defined by mutual induction,
using the grammar below.
Note that messages are defined just as in \secref{s:syntax}.
\begin{eqnarray*}
\msg &::= & t ~|~ k ~|~ n ~|~ i ~|~ (\msg,\msg') ~|~ \{\msg^i\}_k ~|~ \PHI\\
\PHI & ::= & i \believes \PHI ~|~ i \controls \PHI ~|~ i \sees \msg ~|~ 
i \said \msg ~|~ i\key{k}j ~|~ \pkey{k}j  ~|~  
 \fresh(\msg).
\end{eqnarray*}
The superscript $i$ in 
$\encr{\msg^i}{k}$ 
represents a ``from''-field, intended to indicate the 
original sender of the message. 
The intuitive reading of the formulas is as follows:
$i\believes \PHI$ means that  principal $i$ believes formula
$\PHI$; $i\controls \PHI$ means that principal $i$ is an authority on 
or has authority or jurisdiction over
$\PHI$; 
$i \sees \msg$ means that 
$i$ has received a message containing $\msg$;
$i\said \msg$ means
that principal $i$ at some time sent a message containing $\msg$ 
and, if $\msg$ is a formula $\PHI$ that was sent recently, that $i$
believes $\PHI$;  
$\fresh(\msg)$ means that  message $\msg$ was sent recently;
$i\key{k}j$ means that principals $i$ and $j$ can
use the shared key $k$ to communicate (and that the key is a \emph{good} key;
we discuss what counts as a good key below);
$\pkey{k}j$ means that key $k$ is $j$'s public key (and that the
key is a good key).

\begin{figure*}[t]
\hrule
\medskip
\begin{math}
\begin{array}[t]{l}
  \mbox{R1.}\quad
  \Rule{i\believes j\key{k}i\quad i\sees\{\PHI^l\}_k \quad l\ne i}
       {i\believes j\said \PHI} \\
 \mbox{R2.}\quad
  \Rule{i\believes j \said (F,F')}{i \believes j\said F}\\
 \mbox{R3.}\quad
  \Rule{i\believes\fresh(\PHI)\quad i\believes(j\said \PHI)}
       {i\believes j\believes \PHI}\\
 \mbox{R4.}\quad
  \Rule{i\believes j\controls \PHI ~~ i\believes j\believes \PHI}
         {i\believes \PHI} \\
 \mbox{R5.}\quad
 \Rule{i\sees (\PHI,\PHI')}{i\sees \PHI}\\
 \end{array}
 \begin{array}[t]{l}
  
 \mbox{R6.}\quad
  \Rule{i\believes j\key{k}i\quad i\sees\{\PHI^l\}_k \quad l \ne i}{i\sees \PHI}\\
 \mbox{R7.}\quad
  \Rule{i\believes \pkey{k}i\quad i\sees\{\PHI^l\}_k \quad l \ne i}{i\sees \PHI}\\
 \mbox{R8.}\quad
  \Rule{i\believes \fresh(\PHI)}{i\believes \fresh ((\PHI,\PHI'))}\\
 \mbox{R9.}\quad
  \Rule{i \believes i\key{k}j}{i\believes j\key{k}i}\\
\end{array}
\medskip
\end{math}
\hrule
\caption{BAN inference rules}
\label{f:ban}
\end{figure*}
BAN logic uses inference rules to derive new formulas from
others.
These capture the intended meaning of the primitives. 
The most significant rules appear in 
Figure~\ref{f:ban}.  

\subsection{A Probabilistic Interpretation}\label{s:dy}

\begin{tarkin}
We now define a 
translation from BAN formulas to formulas in
our logic.
\end{tarkin}
We write $F^T$ to denote the result  of translating the BAN logic
formula $F$ to a formula in our logic.
Since formulas include messages and are messages, we also need
to translate messages; 
$\mtrans{\msg}$ denotes the translation of a message in the BAN
framework to a message in our framework.
Note that $F^T$ (the translation of $F$ viewed as a formula) is slightly
different from $F^M$ (the translation of $F$ viewed as a message), 
in that the former is of type formula, whereas the latter  is of type
message.   

The translation of messages that are not formulas is
defined inductively in the obvious way:
for a primitive message $\msg$, $\mtrans{\msg} = \msg$, and
$\mtrans{(\msg_1,\msg_2)} = (\mtrans{\msg_1},\mtrans{\msg_2})$.
We translate encryptions $\encr{\msg^i}{k}$ by treating the
``from''-field as concatenated to the end of the encrypted message;
thus, 
$\mtrans{\encr{\msg^i}{k}} = \encr{(\mtrans{\msg},i)}{k}$.
The translation $F^M$ of a formula $F$ viewed as a message
is $\fmla{\ktrans{F}}$, where
$\ktrans{F}$ is the translation of $F$ viewed as a formula.

Our translation for $\believesn$ is based on a definition of belief
due to Moses and Shoham \citeyear{MosesShoham}. 
They assume that an agent operates with a set of default assumptions,
expressed as a formula $A$. An agent's belief in $\phi$, relative to
assumptions $A$, can then be captured by the formula $K_i(A \rimp
\phi)$. 
That is, the agent believes $\phi$ relative to assumptions $A$ if it
knows that $\phi$ holds under assumptions $A$.
\begin{ccsin} 
We use
a probabilistic version of this idea. 
\end{ccsin} 
Like AT, we use a set of good runs
 for the assumptions relative to which the agent reasons. 
Intuitively, these are the runs in which undesirable events such as the
adversary guessing a nonce do not occur.  
We differ from AT in the way that the set of good runs is obtained. 
AT define the good runs by a complicated fixed point construction based
on the original set of beliefs ascribed to the agents by the BAN logic
analysis.  
We allow  any set of runs to be taken as the good runs, but 
typically, the prior probability of the set of good runs would be high 
(a fact that can be expressed in the logic) so that agents have reasonable 
grounds to trust the conclusions they can  draw from the assumption
that a run is good.  
Moreover, for the soundness of one axiom, we need agents to always
assign positive probability to a run being good.

The particular choice of good runs used in proving that a protocol
satisfies a BAN logic specification will depend on the details of the
protocol and the system used to model the behaviour of the adversary. 
Let $\mathit{good}$ be a primitive proposition that expresses ``the
run is good''.  
We take the translation of $i\believes \PHI$ to be 
\begin{ccsin}
$$\neg K_i^0 \neg
\mathit{good} \land K_i^0(\mathit{good}\rimp\ktrans{\PHI})~.$$
\end{ccsin} 
The second clause says that 
$i$ believes $\PHI$ if   \agpr
knows with probability $1$ that when a run is a good,
$\ktrans{\PHI}$ is true.
The interpretation of belief as knowing with probability $1$
is standard in the economics literature. 
We have modified this 
so as to 
make belief depend only on what happens in
the good runs.
The first clause,
$\neg K_i^0 \neg\mathit{good}$,
 requires that the set of good runs has positive 
probability
in at least one cell.
This prevents an agent from vacuously believing a fact
  just
because 
  \agpr knows that 
the probability of a run being good is 0.

The  translation of $\ktrans{(i \sees \msg)}$ is
$K_i(\extract{i}{\mtrans{\msg}})$.  
Thus, agent $i$ ``sees'' $\msg$ if  \agpr knows that  \agpr has extracted it.
We work with this translation here because it helps to satisfy R1, but
it is not the only candidate.  
Another reasonable translation, corresponding to the statement 
that $i$ has received a string that 
he  knows contains the encoding of
$\msg$, is
$\exists x, y(\intn{\mtrans{\msg}} = x\land \recv{i}{y} \land K_i (x
\sqsubseteq y))$.  
The difference lies in whether $i \sees \msg$ is meant to imply that $i$
knows how $\msg$ is composed.

Roughly speaking, BAN interpret 
$i \said \msg$ as ``$\msg$ was a
submessage of a message that $i$ sent at some point in the past''.  The BAN
reading of $\saidn$ also involves claims about belief; BAN
assumes that all formulas said recently by $i$ are believed by $i$.  We
do not make this assumption in our translation, because we do not view
it as an intrinsic part of $\saidn$.   Rather, we capture it in the
systems for which we prove the translation sound.
Given this, we translate $i \said \msg$ as
$$\exists x,y( \intn{\mtrans{\msg}} = x\land \NDiamond \sprev (\neg \send{i}{y} 
\land \Circ \send{i}{y}\land K_i(x \sqsubseteq y)).$$ 
Thus, roughly speaking, $i$ said $\msg$ if at some
point
strictly 
 in the past $i$ sent a string 
 $y=\strin'$, 
 and $i$ knew at the
beginning of the round in which $\strin'$ was sent that 
$x=\intn{\mtrans{\msg}}_r$
was a substring of $\strin'$. 
(There are some significant subtleties in this translation that relate to a
known error in AT identified in  \cite{r:syverson94};
we expand on this in 
Section~\ref{sec:subtleties}.)

Capturing that $k$ is a good key between $i$ and $j$ depends
on what we mean by ``good key''. 
\begin{ccsin} 
There are at least two possible 
interpretations.  One is that no one 
other than possibly $i$ and $j$ has extracted the key.
Accordingly, we 
take $\ktrans{(i\key{k}j)}$ to be $\extract{i}{k}\land\extract{j}{k}\land\bigwedge_{i'\not=i,j}\neg
\extract{i'}{k}. $
This translation would not hold in protocols where the key $k$ is 
provided to $i$ and $j$ by a key server. To cover this, AT
propose the  interpretation ``no one but $i$ and $j$
sends messages encrypted with $k$'' for the length of the protocol
interaction. We could encode this also, as well as other ways to 
make explicit the beliefs of the agents about the behaviour of the key server. 
(See 
Section~\ref{sec:subtleties}
for more discussion of good keys.)
\end{ccsin}

Formula
$\pkey{k}j$ says that $k$ is $j$'s public key,
and that the key is a good key.
The formula is intended to mean that only $j$ knows the key $k^{-1}$.
Thus, its translation is similar in spirit to that of the formula for
shared keys,
and the same comments apply; we take $\ktrans{(\pkey{k}j)}$ to be 
$\extract{j}{k^{-1}}\land \bigwedge_{\{i: i\not=j\}} \neg
\extract{i}{k^{-1}} $.
Of course, situations involving key escrow would require a different
translation. Again, the strength of our approach is that it 
allows us to easily express such variants.

A message is fresh if it could not have been sent, except possibly
recently, where ``recently'' means ``in the last $\tp$ steps''.  We
leave it to the user to decide 
what counts as ``recently'', by choosing a suitable $\tp$. 
Thus, the translation of $\fresh(\msg)$ is
$$\exists x (\intn{\mtrans{\msg}} = x ~\land~
\NCirc^{\tp}\bigwedge_{i}
\NBox (\neg \exists y ( \neg \send{i}{y} \land
\Circ \send{i}{y} \land
 x \sqsubseteq y))).
$$
While it may capture a notion relevant to reasoning about replay attacks, 
this notion of freshness does not capture what we believe should be
meant by a nonce being ``good''.
Intuitively, this is due to the 
requirement for 
unpredictability of the nonce; we return
to this issue in 
Section~\ref{sec:related}.

We interpret  $i \controls \PHI$ 
as ``$i$ believes $\PHI$ if and only if $\PHI$ is true''.
Thus, the translation of $i\controls\PHI$ is
$K_i^0(\mi{good}\rimp\ktrans{\PHI})\riff\ktrans{\PHI}$.
This captures, to some
extent, the intuition that $i$ is an authority on $\PHI$.
Roughly speaking, there is no way for $F$ to change without agent $i$
knowing it, so $F$ is in some sense ``local'' to agent $i$. 

This completes the translation.  For the language that does not
include the $\controlsn$ operator, the translation is linear.  A BAN $F$
formula is translated to a modal formula whose length is linear in $|F|$.
With $\controlsn$ in the language, the translation becomes
exponential.  
It
is not clear that formulas with nested occurrences of
$\controlsn$ arise naturally.  For the language with no 
nested occurrences of $\controlsn$, the translation is again linear.

One other comment: although we have called our translation a
``probabilistic translation'', we in fact use probability in only
a limited way.  The only probabilistic statements that we make are ``with
probability 1'' ($K_i^0$) and ``with probability greater than 0''
($\neg K_i^0 \neg$).  To capture this, we could have simply used a standard
belief operator (that satisfies the axioms of the modal logic KD45),
and avoided dealing with probability 
altogether.  We have used probability here because in the full paper
we consider a more general probabilistic translation, which is also
sound, where we take believing to be knowing with some
probability $\alpha$.  The main consequence of this more general
interpretation is that the translation of the BAN inference rules is
more constrained. 
For example, if a BAN inference rule involves beliefs in the
antecedent and in the conclusion of the rule, the probability
associated with those beliefs must be related.  We leave further
details to the full paper.

\subsection{Evaluating the Interpretation}\label{a:soundness}

To what extent does the translation above capture BAN logic? 
The minimum we can ask for is that the translation validates the
inference rules of BAN logic. 
This is what we argue in this section.

In order to validate the BAN inference rules, we need to 
restrict to systems that satisfy certain properties. Intuitively, these
restrictions are
made implicitly by BAN logic, and must be made explicit in order
to prove the soundness of the translation. 

We say that agents \emph{have no additional prior information
beyond guesses} in an interpreted system $\I$ if the initial
states of all agents includes all public keys, their own
private keys, the nonces required by their protocol (in the
case of nonadversary agents), a finite set of other keys or
nonces they have guessed,
and nothing else.
We also need to 
make precise the intuition that agents tell the truth, since
BAN logic assumes that when a (nonadversary) agent  
sends a formula,  \agpr believes the formula.
Without this requirement, we cannot ensure the validity of 
R3. 
Implicit in the notion of honesty is the idea that an agent
does not forge ``from''-fields in messages. 
Furthermore, BAN logic assumes that 
agents' capabilities of creating and decomposing messages
are those characterized by the Dolev-Yao model.
We capture these capabilities,
together with the assumption that agents not forge ``from''-fields, by 
providing a suitable interpretation of $\extractn$. 
The idea is to define a set of strings $\cancompute{i}(r,m)$, which should
be thought of as the set of strings that $i$ can generate given the
information it has in state $r_i(m)$.  There are two ways that $i$ can
generate a string.  It can pair strings that it can generate to form 
a more complicated string, or it can ``pick apart'' a pair 
into its components.

Suppose that we have a function $\mathit{init}(\st)$ that, given a local
state $\st = \<e_0, e_1, \ldots, e_m\>$, returns the set of strings
contained in
the initial state $e_0$ (roughly speaking, these are the keys and nonces that
$i$ is initially aware of).
Given a point $(r,m)$, 
define $\cancompute{i}(r,m)$ to be the smallest set 
$S$ of 
strings 
satisfying the following conditions:
\begin{enumerate}
\item $\{\strin: \recvE{\strin}\in r_i(m)\} \cup \{\strin: \strin \in
 \mathit{init}(r_i(m))\} 
\cup 
\{j: \mbox{$j$ is an agent}\} 
\cup \{\phi:  \mbox{$\phi$ is a formula}\}
\subseteq S$;
\item $(\strin,\strin') \in S$ iff $\strin, \strin' \in S$;
\item if $\int{\encr{\strin}{k}}{r} \in S$ and $k^{-1} \in
S$, then 
$\strin \in S$ (recall that if $k$ is  symmetric key, then we
identify $k$ and $k^{-1}$);
\item if $\strin, k  \in S$, then
$\int{\encr{\strin}{k}}{r} \in S$. 
\end{enumerate}

We now want to connect $\cancompute{i}(r,m)$ with what $i$ knows about
message terms at the point $(r,m)$.  
We make some additional assumptions regarding what agents know.
Specifically, if $(r,m) \sim_i 
(r',m')$, then we assume  
that (a) if 
$\int{\encr{\strin}{k}}{r}$ and
$k^{-1}$ are both in $\cancompute{i}(r,m)$, then  
$\int{\encr{\strin}{k}}{r} =\int{\encr{\strin}{k}}{r'}$, 
and (b) if $\strin,  k \in 
\cancompute{i}(r,m)$, then $\int{\encr{\strin}{k}}{r} = 
\int{\encr{\strin}{k}}{r'}$.
 Roughly speaking, (a) says that if $i$ ``has'' the
decryption key $k^{-1}$ and ``has'' the string
$\int{\encr{\strin}{k}}{r}$, 
which is the
encryption of $\strin$ under $k$ in $r$, then $i$ knows that 
$\int{\encr{\strin}{k}}{r}$
is the encryption of $string$ 
$\strin$
under $k$, while (b) says that if $i$ ``has''
$s$ and $k$, then  
$i$ knows that $\int{\encr{s}{k}}{r}$ is the encryption
of $\strin$ under $k$.  

With these assumptions, 
as the following result shows, 
$\cancompute{i}(r,m)$ depends only on
$i$'s local state at $(r,m)$.  
\begin{proposition}\label{l:cangenerate}
If $r_i(m) = r'_i(m')$, then 
$\cancompute{i}(r,m) = \cancompute{i}(r',m).$
\end{proposition}

We say interpreted system $\I$ \emph{models agent $i$ as a Dolev-Yao
agent} if
for all $m\ge 0$  and all messages $\msg$, 
\begin{enumerate}
\item 
    $\pi(r(m))(\extract{i}{\msg}) = {\bf true}$ if and only if
$\int{\msg}{r} \in\cancompute{i}(r,m)$, and  
\item 
if $\sendE{i,\strin}\in r_i(m+1)$ and 
$\sendE{i,\strin}\not \in r_i(m)$, 
then
$\strin \in\cancompute{i}(r,m)$. 
\end{enumerate}
Note that the second clause restricts agents 
to sending messages that they can generate.  We
place a further restriction on what are called ``nonforging'' agents.  
Although a nonforging agent $i$ can generate all the messages in
$\cancompute{i}(r,m)$, we assume that, when sending a message, $i$ does not
forge signatures; that is, $i$ will not send a message with a
``from"-field that states that the message is from some other agent.  
Define $\cancomputeNF{i}(r,m)$ just as $\cancompute{i}(r,m)$, except
that condition (4) is replaced by the following variant:
\begin{enumerate}
\item[(4$'$)] 
If $\strin, k  \in S$, then $\int{\encr{\strin,i}{k}}{r} \in S$. 
\end{enumerate}
Rule (4$'$) 
ensures that when the agent constructs an encrypted message,  \agpr
includes a ``from''-field set to its own name.
The definition of a \emph{nonforging Dolev-Yao agent} is 
just like that of a Dolev-Yao agent, except for the use of
$\cancomputeNF{i}(r,m)$ instead of $\cancompute{i}(r,m)$ in the second clause.

Finally, we say that an agent $i$ is honest if, whenever $i$ says
something, then $i$ believes that it will be true when the message is
received.   
(The 
honesty
assumption is how we capture BAN's requirement on $\saidn$
that agents  believe that a formula they are sending is true.)
Formally, agent $i$ is an \emph{honest Dolev-Yao agent} in an
interpreted system $\I$ if
\begin{enumerate}
\item $i$ is a nonforging Dolev-Yao agent in $\I$, and 
\item 
for all BAN formulas $F$, 
\begin{ccsin} 
$$
\begin{array}{l}
\I \sat \exists \strin, \strin' (\intn{\mtrans{F}} = \strin \land \neg
\send{i}{\strin'} \land \Circ
\send{i}{\strin'} \land K_i(\strin \sqsubseteq \strin')) \rimp
K_i^0(\bigwedge_{0 \le \tp'\leq\tp} \Circ^{\tp'}\ktrans{F}).
\end{array}
$$
\end{ccsin}
\end{enumerate} 
The intuition for the last condition is that an
agent says only things that it 
believes 
will still be true some
time in the near future when its message is received.  
Again, this is parameterized by a time $\tp$, which should be taken as
the same time parameter used to interpret freshness.

Observe that while the restriction to Dolev-Yao agents is hardwired
into the definitions of $\saidn$ and $\seesn$ by AT, we model it using
$\extractn$ instead.  This means that our logic can be used to deal
with adversaries other than those that satisfy the Dolev-Yao
properties, without changing the underlying syntax and semantics.
Similarly, rather than hardwiring honesty into the definition of
$\saidn$, we 
model it as an assumption on 
the class of systems.  We can therefore
model the kind of operators BAN advocate without being tied to the
particular choices made by BAN and their successors.

A further assumption we need to make for the soundness of R3 is
regarding our notion of $\good$ runs.  It says that,
in a good run, agents always
consider it possible that runs are good (or, more precisely, assign that
event positive probability). We say that a system $\I$
\emph{maintains 
goodness} if, for all agents $i$ and all points $(r,m)$, we have 
$$(\I,r,m) \sat \good \rimp \neg K_i^0 \neg \good.$$
Note that in any system where all finite prefixes of runs have positive
probability, this axiom will be sound.

We would now like to show that the translation of \secref{s:dy}
preserves the validity of the BAN inference rules.  
Note that an instance of a BAN inference rule has the form
``from $F_1$ [and $F_2$] infer $F_3$''. 
We translate this instance into
a formula of the form $F_1^T [\land F_2^T] \rimp F_3^T$.
Thus, for example, an instance of rule R3 translates to the formula
\begin{ccsin} 
$$
\begin{array}{ll}
(\neg K_i^0 \neg \mathit{good} \land
K_i^0(\mathit{good}\rimp\ktrans{(\fresh(F))})\land 
K_i^0(\mathit{good}\rimp\ktrans{(j\said F)})) \\ \ \ \ \ \rimp
(\neg K_i^0 \neg \mathit{good} \land
K_i^0(\mathit{good}\rimp (\neg K_j^0 \neg \mathit{good} \land
K_j^0(\mathit{good}\rimp\ktrans{\PHI})))).
\end{array}$$
\end{ccsin} 
Note that the translation $\ktrans{(j\said F)}$ involves the message
translation $\mtrans{F}$. 
This is why, for instance, even if $F$ and $F'$ are
equivalent, $\ktrans{(j\said F)}$ and $\ktrans{(j\said F')}$ may not be,
while $\ktrans{(j\believes F)}$ and $\ktrans{(j\believes F')}$ are.

The following theorem, whose proof is in the full paper,
assures us that the translation preserves soundness.
In the theorem statement, we use the notation $r_{ij}^T$ to emphasize
that the formulas in the translation of $r$ refer to agents $i$ and $j$,
\begin{theorem}\label{t:ban1}  
The translation $r_{ij}^T$ of an instance $r_{ij}$ of the BAN
inference rule {\rm R1} is valid in systems that model
Dolev-Yao agents that have no additional prior information beyond
guesses and where
agent $i$ is a nonforging Dolev-Yao 
agent. 
The translation $r_{ij}^T$ of an instance $r_{ij}$ of the BAN
inference rule {\rm R$3$} is valid in systems that model Dolev-Yao
agents that have no additional prior information beyond guesses,
maintain goodness,
 and
where agent $j$ is honest.  
Finally, the translation $r^T$ of an instance $r$ of
{\rm R2} and
{\rm R$n$} for $n\geq 4$ is valid in systems that
model Dolev-Yao agents that have no additional prior information
beyond guesses.
\end{theorem}

Soundness tells us that our translated constructs satisfy properties
analogous to those satisfied by the original BAN constructs.
Of course, there are many translations with this property, some
less interesting than others. 
For example, the translation that sends every BAN formula to the
formula $\truep$ also validates the BAN inference rules.
We hope that the reader agrees that our translation captures the spirit
of the BAN rules.

\subsection{Subtleties in the BAN translation} 
\label{sec:subtleties}

There is an important subtlety 
in the translation of  $i ~\saidn~ \msg$.  
It is actually \emph{not} quite the case that $i$ must know at time
$m'-1$ that 
$\intn{\mtrans{\msg}}$ 
is a substring of $\strin'$; what $i$ must know at
time $m'-1$ is that the string $\strin$ that represents 
$\mtrans{\msg}$ 
in the
current run  is a substring of $\strin'$.  There is a big difference between
$ K_i \exists x ( \intn{\msg} = x \land x \sqsubseteq \strin')$ 
and 
$\exists x (\intn{\msg}  = x \land K_i (x \sqsubseteq \strin'))$.  
A few examples might help to explain the distinction.
First, suppose that, 
in run $r$,
$j$ sends $i$ an unencrypted  string $\strin = \int{\msg}{r}$.  This
means that $i$ can extract $\msg$ in run $r$.  Then $j$ sends $i$ the string
$\strin' = \int{\encr{\msg}{k}}{r}$. 
Finally, $i$ forwards $\strin'$ to some other player $j'$ in round $m'$
of $r$.   
Since $i$ does not have the key $k$ at time $m'-1$, the beginning of the
round  when  \agpr sends $\strin'$
to $j'$, $i$ does not realize that  $\strin'$ represents the encryption of $\msg$.  
That is, although $\strin \sqsubseteq_r \strin'$, there may be a run
$r'$ such that $r'_i(m'-1) = r_i(m'-1)$ but $\int{\encr{\strin}{k}}{r'} \ne
\strin'$.  Thus, $K_i(\strin \sqsubseteq \strin')$ does \emph{not} hold
at $(r,m'-1)$ (although $\strin \sqsubseteq \strin'$ does).
As a consequence, $i \said \msg$ does not hold at $(r,m)$.

Now suppose that $\msg_2 = \encr{\msg_1}{k'}$, 
and that in run $r$,  
$ \int{\msg_2}{r} = \strin$,
where $k'$ is a
key that $i$ does not have, and $j$ sends $i$ 
the string $\strin'=
\int{\encr{\msg_2}{k}}{r}$, 
where $k$ is a shared key between
$i$ and $j$,  
and $i$ then forwards $\strin'$ to $j'$ in round $m'$.  
 Under natural
asssumptions, 
since $i$ has key $k$, 
$i$ ``understands'' that $\strin'$ is the encryption of $\strin$ by $k$, 
so in all runs $r'$ that $i$ considers possible at $(r,m'-1)$, $\strin
\sqsubseteq 
\strin'$.  Thus, 
$(\I,r,m'-1) \sat \intn{\msg_2} = \strin \land  K_i(\strin \sqsubseteq \strin')$, 
so $(\I,r,m'-1) \sat  \exists x ( \intn{\msg_2} = x \land  K_i(x \sqsubseteq \strin'))$.  
On
the other hand, 
$(\I,r,m'-1) \sat \neg K_i\exists x (\intn{\msg_2}= x \land x
\sqsubseteq\strin')$.   
Since $i$ cannot decrypt $\msg_2$, it may well be that
$\int{\msg_2}{r'} \ne \strin$ in some run $r'$ that $i$ considers possible.

As we show, 
our translation of $\saidn$ makes R1 sound; the
alternative translation 
$$\exists y(  \NDiamond \sprev (\neg \send{i}{y} 
\land \Circ \send{i}{y}\land K_i\exists x (\intn{\mtrans{\msg}} = x\land x \sqsubseteq y))$$
does not. 
It is not clear exactly which translation most closely captures what BAN
had in mind for $\saidn$.  We suspect that they were not aware of these
subtleties.  One piece of evidence in support of this suspicion is that,
as previously noted by Syverson and van Oorschot \cite{r:syverson94}
the AT translation of $\saidn$ does not make R1 sound (despite AT's
claims to the contrary).  They run into trouble precisely on examples
like our second example, with nested encryptions.  Note that we are not
claiming that our 
translation is the ``right'' translation of $\saidn$, although something
like our translation seems to be needed to make R1 sound.
We may instead want to consider a different translation, and interpret
$\saidn$ differently. 
What is important is that our logic helps us clarify
the relevant issues.

The meaning of ``good key'' 
in our gloss of the formula $i\key{k}j$ is 
also not as simple as we have made out. 
 Many protocols studied in the literature assume the
existence of a key server in charge of distributing session keys to
principals. In such a context, a good key is not only known to the
principals exchanging messages, but also of course to the server that
initially distributed the key. In some sense, the interpretation of
``good key'' depends on details of the protocol being executed, 
such as 
whether it requires a key server.
This to us suggests that ``good key'' is not an appropriate primitive;
it is a complex notion that is too protocol dependent.  
(Moreover, we would argue that it is better to make explicit 
in the server specification the allowed server behavior with respect to the keys that it generates, 
rather  
than hide this in a primitive of the logic.) 
 In any case,
we can easily accommodate 
such a definition of ``good key'' with trusted servers
by assuming that the server, as well as $i$ and $j$, can extract the key.
For
simplicity, however, we consider only the interpretation of
``good key'' given above. 
Note that both BAN and AT interpret $k$ being a good key as a 
statement that also talks about the future; in essence, if $k$ is
a good key, it remains so throughout a protocol interaction. We can
capture such an interpretation by prefixing the translated formula by
a $\Box$ operator. 
Of course,
this interpretation precludes the analysis
of protocols that leak the key value. (See Nessett \citeyear{r:nessett90}
and Burrows, Abadi and Needham \citeyear{r:burrows90a}
for discussion.) We would like the analysis to reveal such leaks, rather
than presupposing that they do not happen.

There is yet another subtlety.  It is consistent with our translation
that $k$ is a good key between $i$ and $j$, but neither $i$ nor $j$
knows this.  One obvious reason is that $i$ and $j$ may consider it
possible that the key has leaked.  We might then hope that $i$ and $j$
know that, if the run is $\good$ (so that there is no leakage), then $k$
is a good key.  But even this may not be the case under our definition.  For
example, $i$ may not know that $j$ can extract $k$.  Although we do not
need it to show that the BAN axioms are sound, we might consider
requiring that, on runs that are $\good$, $i$ and $j$ know that 
a key shared between them is good.  More precisely, let $E^{\good}_G \phi$
be an abbrevation $\land_{i \in G} K_i(\good \rimp \phi)$; that is,
all agents in $G$ know that, if the run is good, then $\phi$ holds.  
We might want to require that $\ktrans{(i\key{k}j)}$ implies
$E^{\good}_{\{i,j\}} \phi$ (where $\phi$ is the translation we give
above).  

We could go even further.  Rather than just requiring $i$ and $j$ to
know that if the run is good, the key is shared, we might want this fact
to be common knowledge among $i$ and $j$.  To make this precise, 
let $(E^{\good}_G)^{h+1}\phi$ be an abbreviation
   for $E^{\good}_G ((E^{\good}_G)^h \phi)$.
Define $C^{\good}_{G} \phi$ so that it  holds exactly if 
$(E^{\good}_G)^h \phi$ 
holds for all $h
\ge 1$.
We could then consider taking $\ktrans{(i\key{k}j)}$  to be
$C^{\good}_{\{i,j\}} \phi$, where $\phi$ is the translation we used
above.  We did not do this, in part because it is not clear that it is
really desirable to make such strong requirements for shared keys.  For
example,  
suppose that in an environment where message transmission is not
completely reliable (so messages may be lost), $i$ and $j$ already have
a shared  key $k$, and $i$ decides that it should be refreshed.  So $i$
tells $j$ that $k$ should be replaced by $k'$ (using a message encrypted
by $k$).  While $j$ can acknowledge receipt of the message and $i$ can
acknowledge the acknowledgment, as is well known, 
no amount of back and forth will make it common knowledge that $k'$ is a
shared key, even if all runs are good \cite{HM90}.  Nevertheless,
this should not prevent $i$ and $j$ from starting to use key $k'$ as a
shared key.
Again, the main point we want to make here is not that our translation
is the ``right'' translation (that will typically be
application-dependent),  but that our logic lets us clarify the issues.

\section{Related Work} 
\label{sec:related}

The goal of understanding the foundations of 
authentication 
logic is not new, and
goes back to early attempts at providing a semantics for the original
BAN logic. 
As we mentioned, BAN logic was originally defined through
a set of inference rules without a semantics tied to the actual
protocols that the logic was meant to analyze.  
The work of Abadi and Tuttle\citeyear{AT91} and others
\cite{r:gong90,r:syverson94,r:stubblebine96,r:wedel96} sought to
provide a direct semantics for both BAN logic and its subsequent
generalizations meant to make it more widely applicable. 
This is not the place to trace the history of BAN logic, but 
a big point of contention has always been
the idealization needed to be performed on the
protocols~\cite{r:mao95}.
Idealization can be understood as a way to ascribe a ``meaning'' to
the various steps of a protocol, and some work has gone towards
understanding such idealization~\cite{r:mao95,r:kindred97}. 
In contrast to providing a direct semantics and an account of
idealization, we instead supply a semantics to BAN logic operators by
decomposing them into more primitive and well-understood logical
operators; 
our semantics makes it possible to avoid
idealization altogether by analyzing the
multiagent systems generated by the protocol under consideration, 
subject to an abstraction of cryptography that captures that agents have 
uncertainty about how cryptography works. 

Several logics for reasoning about security protocols based on
knowledge and not subject to idealization have been proposed, going as
far back as CKT5 \cite{r:bieber90}. 
Van der Meyden 
and Su have used epistemic logic to
model check the dining cryptographers protocols~\cite{r:meyden04}. 
In many such logics (e.g., \cite{r:accorsi01,r:toninho10}) knowledge or
belief is not interpreted as truth at all possible worlds, but rather
as a form of algorithmic knowledge~\cite{r:halpern02e}, much as with
our $\extractn$ primitive, but with a fixed semantics.
Surveys of the application of epistemic logic to reason about 
security protocols include \cite{DechesneWang2010,Pucella2015}. 
In general, work in this area does not use probabilistic  operators, as we have done: 
we think that ultimately, security needs to be analysed probabilistically.

Cohen and Dam~\cite{cohenthesis,r:cohen05,r:cohen07} identified some
of the same subtleties
we identified in the interpretation of the BAN logic operators, but 
they address those issues differently. 
They develop two different semantics that use ideas from counterpart theory 
and the  \emph{de dicto}/\emph{de re} distinction 
in modal logic to address the logical omniscience problem. 
Similar to AT, what is sent and received in the semantics are 
 \emph{message terms}, rather than \emph{strings}, as in our work. 
Both semantics work with permutations $\rho$ on the set of message terms, 
which also extend to transformations on local states. 
The first semantics \cite{r:cohen05} defines knowledge at a global state $s$ as 
$s \models K_i\phi$ if for all global states $s'$ and permutations $\rho$ such that $s'_i = \rho(s_i)$, we have $s' \models \rho(\phi)$. 
This semantics validates the BAN style axiom 
$\recv{i}{m} \rimp K_i \recv{i}{m}$ for all messages $m$, including
messages $m$ such as $\encr{M}{k}$ in situations where $i$ 
cannot extract $k$. 
The second semantics \cite{r:cohen07}, which, judging from \cite{cohenthesis},
they appear to consider to be the more satisfactory,  
makes a distinction between semantic message variables $x$ and syntactic message variables $m$, 
and allows quantification over both. Here the semantics adds an assignment $V$ from 
 variables to equivalence classes of 
 ground message 
 terms with respect to equations capturing 
 the behaviour of cryptography,  such as $\encr{\encr{M}{k}}{k} = M$. They define 
\begin{itemize} 
\item $s,V \models K_i\phi$ if for all global states $s'$ and permutations $\rho$ such that $s'_i = \rho(s_i)$, we have 
$s',\rho \circ V \models \phi$.
\item $s,V \models \forall x (\phi)$ if for all equivalence classes $e$ of ground message terms,   we have $s,V[x\mapsto e]  \models \phi$.
\item $s,V \models \forall m (\phi[m/x])$  if for all ground message terms $M$,   we have $s,V \models \phi[M/x]$.  
\end{itemize}  
This semantics validates  the formula  $\forall x(\recv{i}{x} \rimp K_i \recv{i}{x})$ but not the formula 
$\forall m(\recv{i}{m} \rimp K_i \recv{i}{m})$. The latter is equivalent to 
the conjunction of $\recv{i}{M} \rimp K_i \recv{i}{M}$ for all messages $M$. 
In effect, this semantics treats semantic message variables $x$ somewhat similarly to 
our \emph{string} interpretation of messages, but whereas our 
logic treats message terms and strings as having distinct types, 
they allow formulas such as $x=M$ where a semantic variable $x$ is equated
to a message term $M$. The semantics of our logic more concretely matches
an operational semantics, and maintains a type distinction between 
messages and strings, so we would express the same equivalence as 
$x=[M]$. It would be interesting to investigate whether there is 
any formal correspondence between the two approaches. (One apparent
obstacle to this is that our semantics allows a situation where an agent considers
it possible that a string it has received is the encoding of \emph{no} term
(e.g., it is a random string injected in a guessing attack by the adversary) 
whereas in the Cohen and Dam semantics, everything that is received
is 
semantically an equivalence class of message terms.)

Our approach to logical omniscience is similar to models used in the
cryptographic literature in which encryption is modelled as a
randomly chosen function, not known to the agents
\cite{GoldreichGM86},  
who discover its behaviour by making calls to the function on particular values. 
Such models can be captured in 
our
framework by appropriate construction of the interpreted system. 
(It is more common in cryptography, however, to use the \emph{random oracle} model \cite{CanettiGH04}, 
in which the random function represents a hash function rather than an encryption 
function, and to use this as the basis for the construction of ciphers.)   
Van Ditmarsch et al \cite{Ditmarsch12} use a similar idea, but their approach is purely epistemic and 
they do not have probability in their models. The connections that we draw to BAN logic are not 
developed in this work.

We are careful in our interpretation to distinguish between
the freshness of a nonce---that it has not been used before---from its
unpredictability---how likely it is to be guessed. 
Freshness is captured by the simple statement that no message containing
the nonce was recently sent, while unpredictability is captured by a
probability distribution over the choice of nonces during a run of the
protocol. 
Thus, these two aspects of nonces are kept separate. 
There has been little discussion in the literature about this
distinction, and nonces are often implicitly taken to be unpredictable, 
even though the framework of analysis used technically captures only
freshness (for instance, the spi calculus~\cite{r:gordon99}, or
MSR~\cite{r:cervesato99}). 
When not required to be unpredictable, nonces can be taken to be
sequence numbers, or timestamps.
There has been some work 
discussing
the use of timestamps for nonces
(e.g., Neuman and 
Stubblebine 
\citeyear{r:neuman93}). 
Somewhat related are the notions of authentication tests described by
Guttman \citeyear{r:guttman02a},
where a distinction is made between using 
nonces that, from the perspective of a given agent, should be secret
when sent (and therefore should be unpredictable) from those that need
to be secret only when received.

The probabilistic interpretation of BAN logic tells us that the
symbolic reasoning done in BAN logic can be understood in terms of more
realistic probabilistic beliefs, accounting for the probabilities
associated with the choice of nonces 
and the choices of keys.
{F}rom that perspective, our work 
resembles 
a general class of
work that seeks to obtain results about more realistic models of
cryptography, either by using symbolic reasoning and showing that
results of such an analysis can be interpreted quantitatively, or by 
using a logic that is explicitly more quantitative.

Abadi and Rogaway~\citeyear{r:abadi02a}, building on previous work by
Bellare and Rogaway \citeyear{r:bellare93}, compare the results
obtained by a symbolic analysis with those obtained by a more
computational view of cryptography. 
They show that, under various conditions, the former is sound with
respect to the latter, that is, terms that are assumed
indistinguishable in a symbolic analysis remain indistinguishable
under a concrete encryption scheme. 
This work has been extended in various ways (e.g.,
\cite{r:micciancio04}).

Other formal approaches to security protocol analysis have tried to
apply techniques from symbolic analysis  directly to more realistic
models of cryptography (such as computational
cryptography~\cite{r:goldreich01}) by viewing messages as strings of
bits and adversaries as randomized polynomial-time algorithms. 
The resulting logics are fundamentally probabilistic, in a way similar
to our probability-based interpretation of BAN logic. 
Examples of such approaches include those of Backes, Pfitzmann, and
Waidner~\citeyear{r:backes03} and Datta et al.~\citeyear{r:datta05}, as
well as automated tools such as CryptoVerif~\cite{r:blanchet08}. 
In general, it is difficult in such models to express local
information, that is, the 
fact
that at a given point in the
protocol, a particular agent has certain information.
These approaches are geared instead
towards more global properties, which say 
there is some probability of a particular formula holding at all points in the execution (or at the end of the interaction).

\section{Conclusion}

We have introduced in this paper a simple modal propositional logic to
reason about security protocols, based on well-understood modal
operators for knowledge, probability, and time. We have shown how
those primitive notions can be used to capture the more high-level
operators of BAN logic, helping us to understand the intuitions
underlying BAN logic
and, more importantly, to capture important aspects of reasoning about
security protocols that we claim cannot be adequately expressed in a
non-epistemic way.
A further advantage of the translation is that it allows us to apply
well developed model-checking techniques 
\cite{BoureanuCL09,r:gammie04,HuangLM11,LomuscioQR17,r:meyden04} 
to verifying the
correctness of security protocols whose specifications are expressed
using BAN logic.  (Here it becomes
useful that the translation is linear, at least, if we restrict the
use of the $\controlsn$ operator.)

Our ultimate goal is to design a logic that will be useful for
reasoning about all aspects of security, based on the logic we
introduced here. In this paper, we have focused on high-level issues
concerning the expressibility of high-level operators using
well-understood primitive concepts. Other aspects of reasoning about
security protocols that we have hinted at need to be further
investigated. 
A particularly significant aspect is how to deal with the issue of what
an adversary can 
compute; in this paper, we have sidestepped the problem using the
proposition $\extractn$. 
We are currently investigating more general approaches to deal with
this issue.

\section*{Acknowledgements} We thank the TARK reviewers for their
insightful comments.
Halpern was supported in part by NSF grants
IIS-0911036 and CCF-1214844, by ARO grant W911NF-14-1-0017, 
and by the Multidisciplinary
University Research Initiative (MURI) program administered by the
AFOSR under grant FA9550-12-1-0040.
Van der Meyden was supported by  ARC grant DP120102489.

\bibliographystyle{eptcs}
\bibliography{riccardo2,z,joe,ron}

\end{document}